\documentclass[twocolumn,showpacs,amsmath,amssymb,aps,pra]{revtex4}
\usepackage{graphicx,latexsym}
\pdfoutput=1
\begin{document}

\title{Fast production of  large $^{23}$Na Bose-Einstein condensates in an optically plugged magnetic quadrupole trap}

\author{Myoung-Sun Heo}
\author{Jae-yoon Choi}
\author{Yong-il Shin}\email{yishin@snu.ac.kr}
\affiliation{Department of Physics and Astronomy, Seoul National University, Seoul 151-747, Korea}

\date{\today}

\begin{abstract}
We demonstrate a fast production of large $^{23}$Na Bose-Einstein condensates in an optically plugged, magnetic quadrupole trap. A single global minimum of the trapping potential is generated by slightly displacing the plug beam from the center of the quadrupole field. With a dark magneto-optical trap and a simple rf evaporation, our system produces a condensate with $N\approx10^7$ atoms every 17~s. The Majorana loss rates and the resultant heating rates for various temperatures are measured with and without plugging. The average energy of a spin-flipped atom is almost linearly proportional to temperature and determined to be about 60\% of the average energy of a trapped atom. We present a numerical study of the evaporation dynamics in a plugged linear trap.
\end{abstract}

\maketitle

\section{Introduction}
Ultracold quantum gases have presented a new class of systems to investigate many-particle quantum phenomena~\cite{BlochRMP80,LewensteinAdvPhys56,JakschAnnPhys315}. Since ultracold gas samples are very dilute at particle densities around $10^{13}$ or $10^{14}$/cm$^3$, we need extremely low nanokelvin temperatures to reach quantum degenerate regime. Evaporative cooling has been the powerful and exclusive method for producing quantum gas samples such as Bose-Einstein condensates (BECs)~\cite{Anderson95BEC,Davis95BEC,BarrettPRL87}. Rapid and efficient evaporation requires fast rethermalization and conservative environment minimizing undesired atom loss and heating due to, e.g., inelastic collisions and background collisions~\cite{KetterleAAMOP37}. In this paper, we describe a fast evaporation to produce a large condensate in an optically plugged magnetic quadrupole trap.

A quadrupole trap provides a large trap volume and tight confinement due to the linearity of the potential, which make it beneficial to efficient evaporation. Also the quadruple field cab be generated from a simple pair of coaxial coils with opposite currents, allowing large optical access to a sample. However, Majorana spin-flip loss is a significant drawback in the quadrupole trap. Near the zero-field point at the trap center, nonadiabatic spin transitions to an untrappable state cause atom loss and resultant heating, eventually limiting the maximum gain in phase-space density~\cite{ketterle95}. Several techniques have been implemented to avoid this detrimental effect, including applying a time-orbiting bias field~\cite{Anderson95BEC} or an repulsive optical plug potential~\cite{Davis95BEC,Naik2005BEC}, or transferring to an optical dipole trap~\cite{Lin2009BEC}.

The optical plugging scheme preserves the linearity of the trapping potential, so it is advantageous for fast evaporation. The first $^{23}$Na BEC of $\approx5\times10^5$~atoms was produced in 7~s evaporation~\cite{Davis95BEC}. Recently, high efficiency of evaporation cooling in a plugged trap has been demonstrated by producing a large condensate of $\approx3.0\times10^7$~atoms in 40~s evaporation~\cite{Naik2005BEC}. Here we report a rapid production of a large condensate with $N\approx1.0\times10^7$~atoms in 14~s evaporation. The distinct feature of our system is that the plug beam is offsetted from the magnetic zero-field point. This seems contradicting to the role of the plug, but we find that a slight misalignment less than a beam waist is favorable for BEC production because it forms a single global minimum in the combined magnetic and optical potential. The pointing stability of the plug beam was improved by using a fiber, which makes the fine adjustment of the plug position possible.

Furthermore, we have investigated the effects of the Majorana loss and the plug by measuring the loss rates and the resultant heating rates for various conditions. The average loss energy of a spin-flipped atom shows an almost linear dependence on temperature and is determined to be about 60\% of the average energy of a trapped atom. For a plugged trap, the reduction factor in the loss rate is found to be exponentially proportional to a Boltzmann factor. This confirms that the plugging effect results from density suppression due to the repulsive optical potential in the loss region. Finally, we present numerical simulation for quantitative description of the evaporation dynamics in a plugged quadrupole trap.

\section{Experiment}

\begin{figure}[t]
  \includegraphics[scale=1]{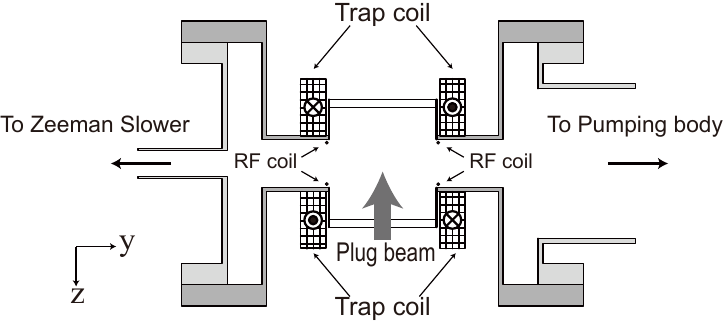}
  \caption{Experimental apparatus. Recessed bucket windows with 60~mm diameter clearance are mounted above and below the main vacuum chamber. A pair of 36 turn water-cooled coils are secured around the windows, generating a magnetic quadrupole field at the chamber center.  Radio-frequency (rf) evaporation coils are placed inside the vacuum chamber. A plug beam is aligned with the axial direction of the magnetic field.}\label{fig:mainchamber}
\end{figure}

Our BEC apparatus employs a zero crossing Zeeman slower, designed to generate a flux of about $10^{11}$ atoms/s~\cite{MITmachine}. Slowed atoms were collected for 3~s in a dark magneto-optical trap (MOT)~\cite{darkMOT} and the atoms in the hyperfine state $|F=1,m_F=-1\rangle$ were transferred in a magnetic quadrupole trap, where $F$ is the total angular momentum and $m_F$ is the magnetic quantum number. The magnetic field was generated from a pair of 36 turn coils, secured around recessed bucket windows of our main experimental vacuum chamber [Fig.~\ref{fig:mainchamber}]. The coils are made from 1/8 in. hollow square copper tubes and cooled with water flowing through the tubes. The average coil diameter is 90~mm and their spacing is about 63~mm. We measured a gradient of 1.5~G/cm/A along the symmetry $z$-axis. For the MOT and the magnetic trap, we applied a field gradient $B'=7.5$~G/cm and 180~G/cm with 5~A and 120~A in the trap coils, respectively. The atom transferring was simply done by ramping $B'$ at a rate of about 10~G/cm/ms and simultaneously turning off the MOT laser beams. The initial number and temperature of the atoms in the magnetic trap were $N_i\simeq3.0\times10^9$ atoms and $T_i\simeq250~\mu$K, respectively.

Optical plugging was performed by focusing a blue-detuned, 532~nm laser beam near the zero of the magnetic field. The plug beam propagated along the $z$-direction with a beam waist of $w\simeq 45~\mu$m, providing a repulsive potential barrier of $U_p\simeq k_B \times 90~\mu$K with 1.7~W, where $k_B$ is the Boltzmann constant. In order to improve its pointing stability, we used a fiber to deliver the plug laser beam close to the main chamber, which was critically important for our BEC production~\cite{footnote1}. For several weeks of operation, the plug position needed little or no adjustment. Finally, rf evaporative cooling was applied for 14~s to the atoms in the plugged quadrupole trap.

\subsection{Quadrupole trap with off-centered plug beam}

\begin{figure}[t]
  \includegraphics[scale=1.0]{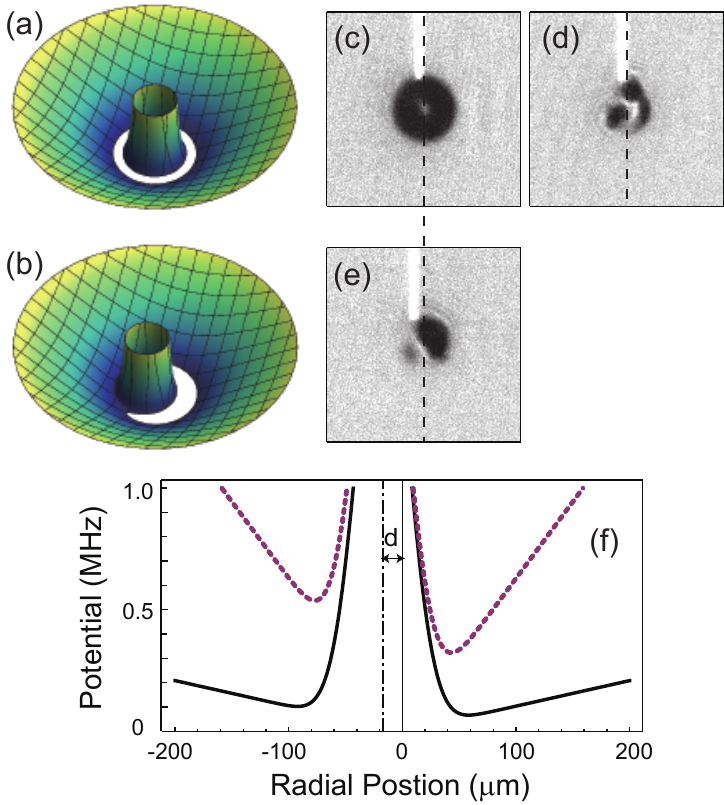}
  \caption{(Color online) Off-centered plugged trap. Schematic diagrams of the combined trapping potential with a plug beam (a) passing the field center and (b) slightly displaced from the center. The ring symmetry is broken by the displacement and a single minimum is generated in the trap. In-situ absorption images of atoms in a ring trap after evaporation with a final rf of (c) 150~kHz and (d) 110~kHz, where multiple small condensates containing $<10^6$ atoms are produced due to the imperfection of the plug beam profile. (e) In-situ image with displacing the plug beam by $d\approx 17~\mu$m and a final rf of 105~kHz. A condensate of $\approx10^7$ atoms is produced in a single potential minimum. The white trace lines in the images are from the leakage plug beam out of filters, which are used for determining the relative beam position. The field of view is 750~$\mu$m$\times750~\mu$m. (f) Radial potential along the trap minimum at $B'=30$~G/cm (solid) and 180~G/cm (dashed). }\label{fig:plug}
\end{figure}

When a circular plug beam is placed at the center of the quadrupole field, the combined potential has azimuthal symmetry, forming a ring trap [Fig.\ref{fig:plug}(a)]. This ring geometry is interesting in terms of persistent current and vortex generation~\cite{Ryu2007Ring}. However, it is desirable to break this symmetry and generate a single point minimum in the trapping potential for controlled production of condensates. Furthermore, the critical temperature in a ring trap for a given atom number is typically low because of no confinement along the azimuthal direction. In previous experiments, this ring symmetry was broken by applying a plug beam in the transverse direction~\cite{Davis95BEC} or using an elliptical beam~\cite{Naik2005BEC}. In our experiment, we simply displaced the plug beam from the zero of the magnetic field by $\simeq17~\mu$m [Fig.~\ref{fig:plug}(b)]. It has been emphasized that misalignment of a plug is detrimental to BEC production~\cite{Naik2005BEC}. We found that a slight misalignment smaller than a beam waist is effectively helpful without significantly losing the plugging effect. The position stability of our plug beam was estimated to be less than 5~$\mu$m.

The effective potential for atoms in the plugged trap is
\begin{equation}
U(\textbf{r})=\mu B'\sqrt{\frac{x^2}{4}+\frac{y^2}{4}+z^2}+ U_{p}\exp (-\frac{(x+d)^2}{2w^2})- mgz,
\end{equation}
where $\mu=\mu_B/2$ ($\mu_B$ is the Bohr magneton) and $m$ are the magnetic moment and mass of the atom, respectively, $d$ is the plug beam offset from the field center, and $g$ is the gravity acceleration. We have ignored the beam focusing because the Rayleigh length $\pi w^2/\lambda\approx12$~mm is larger than the sample size. For our typical final condition with $B'=30$~G/cm and $d=17~\mu$m, we estimate the trap minimum to be located at $(x_0,y_0,z_0)=(60,0,15)~\mu$m with a magnetic field of about 100~mG. From the potential local curvature, the trapping frequencies are estimated to be $(f_x,f_y,f_z)\approx(210,55,110)~$Hz. With $d=0.4w$, the plugging potential at the zero-field point is reduced by about $25\%$ to $k_B\times68~\mu$K. At low temperatures, we observed multiple atom clouds due to small imperfections in the profile of our plug beam [Fig.~\ref{fig:plug}(d)]. The potential variation leading to this fragmentation was determined to be about $h\times$10~kHz from the density distributions for various final rf values, where $h$ is the Planck constant. In the off-centered plugged trap, a single potential minimum was formed at $h\times30$~kHz lower than next potential minimum [Fig.~\ref{fig:plug}(e)].

\subsection{Evaporation process}

\begin{figure}[t]
  \includegraphics[scale=1.0]{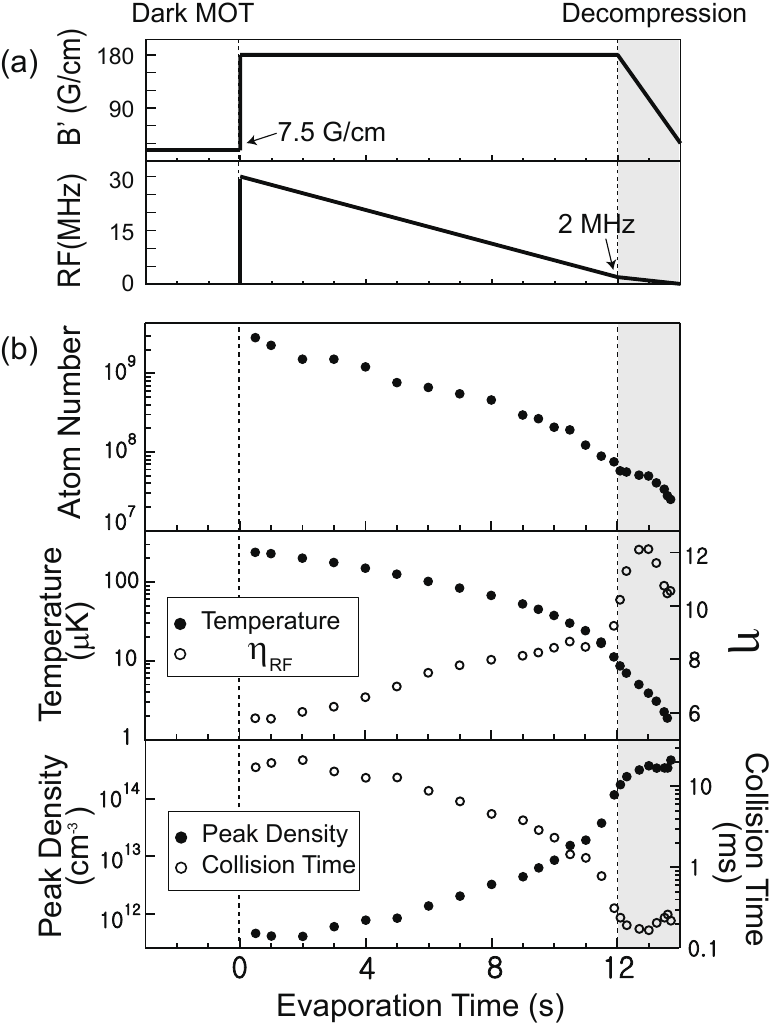}
  \caption{Experimental sequence for BEC production consists of dark MOT, rf evaporation and trap decompression. (a) Field gradient and radio-frequency. (b) Time evolution of atom number, temperature, truncation parameter $\eta$, peak density, and collision time. The decompression phase is marked with a grey zone. }\label{fig:evap}
\end{figure}

Evaporative cooling was performed by linearly sweeping the applied radio frequency $\nu_{\textrm{rf}}$ from 30~MHz to 2~MHz in 12~s and then to 0.1~MHz in another 2~s [Fig.~\ref{fig:evap}(a)]. The quadrupole field gradient $B'$ was ramped down from 180~G/cm to 30~G/cm in the last 2~s. This trap decompression was necessary to achieve a large condensate, reducing inelastic losses associated with high atom density such as three-body recombinations. Furthermore, decompression results in adiabatic cooling of atom cloud, thus relatively enhancing the plugging effect. During the adiabatic process in a linear trap, atom density and temperature scale as $B'$ and $B'^{2/3}$, respectively.

The time evolution of atom number and temperature during the evaporation was determined from absorption images taken after 18~ms time-of-flight [Fig.~\ref{fig:evap}]. Peak density $n$ and collision time $\tau$ are estimated as $n=N/(32\pi (k_B T/\mu B')^3)$ and $\tau^{-1} =\sqrt{2} n \sigma \bar{v}$, respectively, ignoring gravity and the plug potential. $\sigma=8\pi a^2$ is the elastic collision cross section with scattering length $a=52 a_0$ ($a_0$ is the Bohr radius)~\cite{nistsodium} and $\bar{v}= (8 k_B T/ \pi m)^{1/2}$ is the average thermal velocity. The truncation parameter $\eta$ is defined as $\eta=(h \nu_\textrm{rf} - U_0)/k_B T$, where $U_0$ is the potential at the trap bottom, $U_0/h\approx320$~kHz at 180~G/cm and $\approx70$~kHz at 30~G/cm [Fig.~\ref{fig:plug}(f)]. Runaway evaporation was achieved with increasing collision rate $\tau^{-1}$. When the decompression started, the peak density was estimated to be $\sim10^{14}$/cm$^3$. $\eta$ smoothly increased from 6 to 9 before the decompression.

\begin{figure*}[t]
  \includegraphics[scale=1]{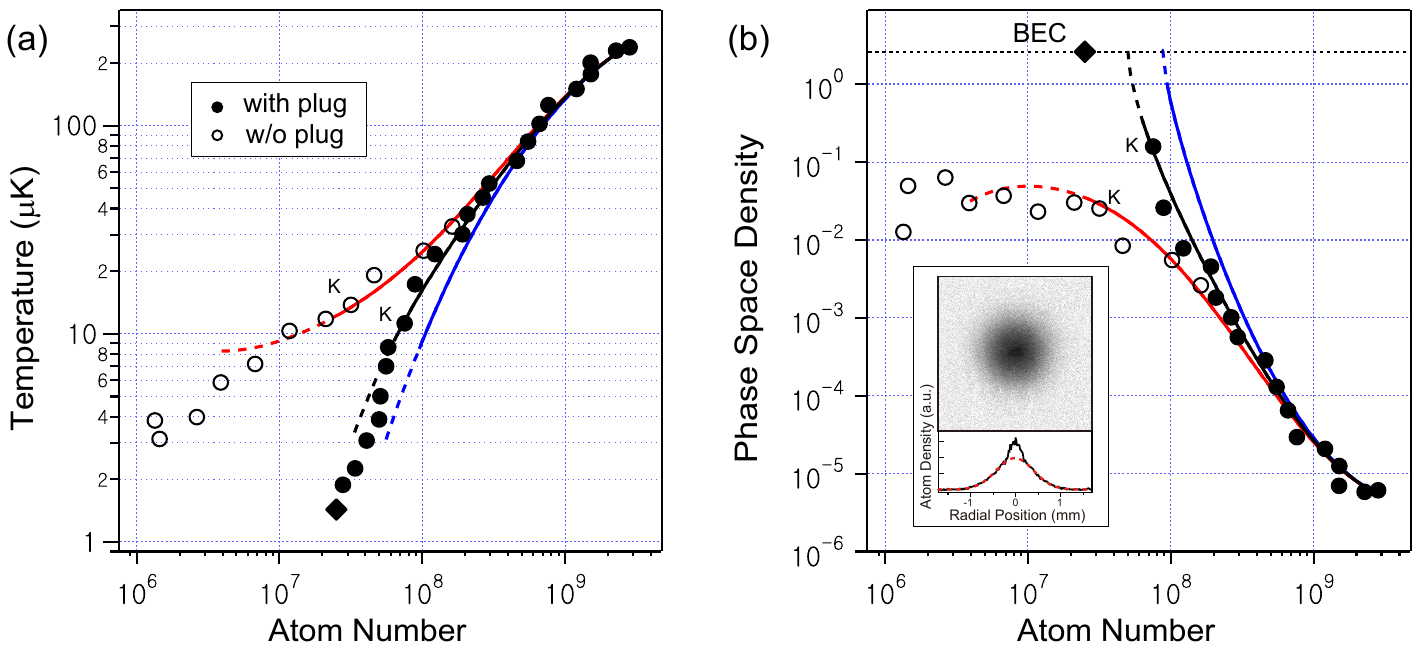}
  \caption{(Color online) Evaporation trajectory. (a) Temperature versus atom number. (b) Peak phase space density versus atom number. Data points at 11.9~s (before decompression) are marked with `K'. The onset of condensation is observed at 13.7~s in a plugged trap from the bimodal momentum distribution (see inset). The black diamond indicates the transition point. The solid lines are from numerical simulation with a plug (black), no plug (red), and no loss (blue). The dashed lines are the evolution after 12~s (see Sec.~\ref{sec:numeric}).}\label{fig:curve}
\end{figure*}

Bose-Einstein condensation was observed after 13.7~s evaporation from the onset of bimodal distribution in absorption image [Fig.~\ref{fig:curve}(b)]. At this transition point, the atom number and the temperature were $N_c=2.5\times10^7$ and $T_c=1.4~\mu$K, respectively. The overall evaporation efficiency is $\gamma=-\ln(D_c/D_i)/\ln(N_c/N_i)=2.7$, where $D$ is peak phase space density given as $D=n \lambda_{dB}^3$ ($\lambda_{dB}=h/\sqrt{2\pi m k_B T}$ is the thermal de Broglie wavelength) and the critical phase space density is $D_c=2.61$. With 14~s full evaporation, an almost pure condensate with $N=1.0\times 10^7$ was obtained. After an additional decompression to $B'=15$~G/cm, the lifetime of the condensate was 10~s, which is long enough for further manipulation such as transferring into an optical dipole trap.

\begin{figure}[t]
  \includegraphics[scale=1]{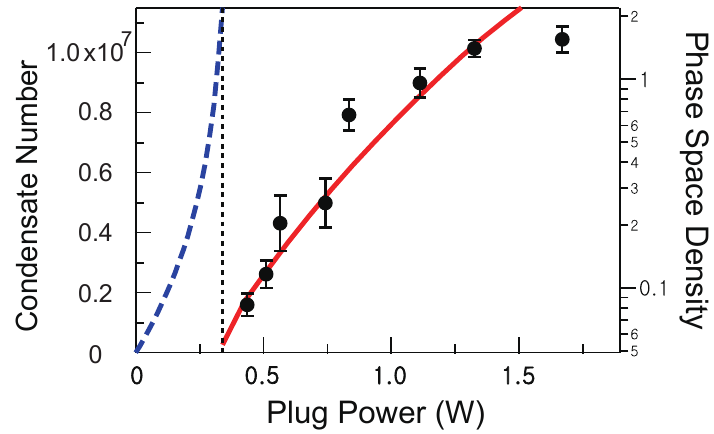}\\
  \caption{(Color online) Condensate number versus plug power. Each point is averaged over three separate runs of the experiment. Maximum phase space density (blue dashed) and condensate number (red) from numerical simulation (see Sec.~\ref{sec:numeric}).}\label{fig:power}
\end{figure}

The evaporation in a bare quadrupole trap was also investigated without the plug beam. Figure~\ref{fig:curve} shows the evaporation trajectories in the $N$-$T$ and $N$-$D$ planes. At $T<20~\mu$K, the plugging effect becomes noticeable from the deviation between the evaporation curves with and without plugging. The maximum phase space density was about $5\times10^{-2}$ in a unplugged quadrupole trap. The minimum plug beam power for making a condensate of $>10^6$ atoms was $\sim0.4$~W for the 14~s evaporation [Fig.~\ref{fig:power}]. In our experiment setup, we could minimize the evaporation time to 8~s to observe a condensate.

\subsection{Effects of Majorana loss}
\begin{figure}[t]
  \includegraphics[scale=1]{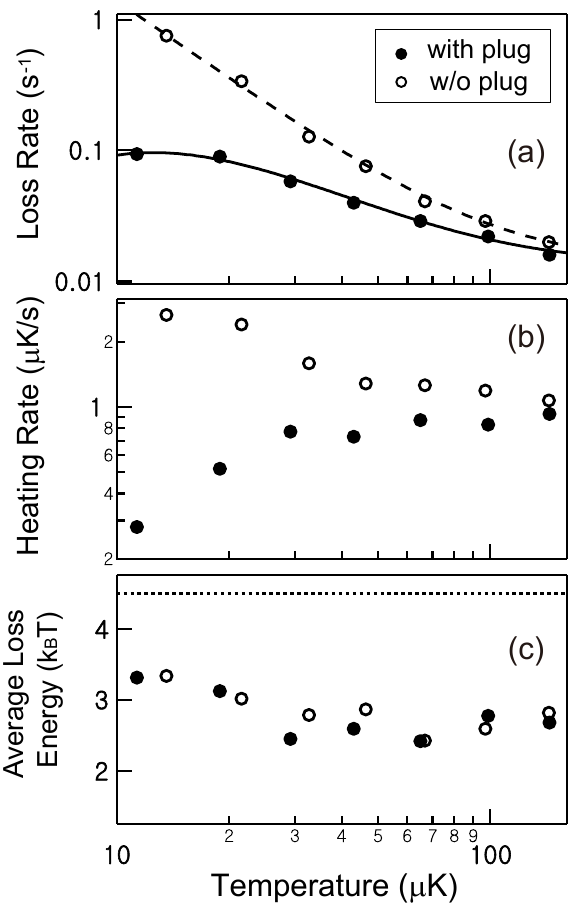}\\
  \caption{Majorana loss in a magnetic quadrupole trap. (a) Loss rates and (b) heating rates for various temperature at the field gradient $B'$=180~G/cm. Solid curves are fits to the data with a loss model, Eq.~(\ref{eq:gamma}) (see text for detail). (c) Average energy of a lost atom due to the Majorana spin-flip. The dot line indicates the average energy of an atom in a linear trap.}\label{fig:Majorana_theory}
\end{figure}

The Majorana spin flips cause heating as well as atom number reduction. To quantitatively understand the loss effects, we have measured the loss and heating rate of a thermal gas for various temperatures with and without the plug at $B'$=180~G/cm [Fig.~\ref{fig:Majorana_theory}]. The initial sample temperature was controlled with the final value of radio frequency. The time evolution of atom number and temperature were determined with various holding times. The rf field was turned off after the sample preparation in order to exclude the evaporation effects~\cite{footnote2}. We observed that the temperature almost linearly increases with a small acceleration and the number decay rate decreases slowly. The initial loss and heating rate were determined from an exponential and a linear fit, respectively, to the data points with temperatures lower than 120~\% of the initial temperature.

A simple argument in Ref.~\cite{PetrichPRL74} suggests that the Majorana loss rate $\Gamma_m$ in a quadrupole trap scales as $\Gamma_m \propto \hbar/ m (\mu B'/k_B T)^2$. A fit to our data in Fig.~\ref{fig:Majorana_theory}(a) with $CT^{-2}+\Gamma_b$ yields $C=139 (3)~\mu$K$^2$/s and $\Gamma_b=0.013(7)$~s$^{-1}$, where $\Gamma_b$ is the background loss rate (corresponding to the lifetime of 75~s). This result suggests $\Gamma_m=0.14~\hbar/ m (\mu B'/k_B T)^2$. The proportionality constant is found to be much smaller than the values of 0.58~\cite{PetrichPRL74,footnote3} and 0.87~\cite{Lin2009BEC} in previous experiments with $^{87}$Rb atoms, which is beyond the simple model. Note that the lifetime of $^{23}$Na atoms in a quadrupole trap is comparable to that of $^{87}$Rb atoms even with their mass difference.

A spin flip happens when the spin of a moving atom cannot adiabatically follow the field orientation, so the flipping probability of an atom at a given position is fully determined by the local field strength, its gradient, and the velocity distribution of the atom, i.e. temperature. Therefore, the plugging effect is a consequence of the density suppression due to the optical repulsive potential in the low-field, spin-flipping region. Including a Boltzmann density suppression factor, the reduced loss rate would be given as
\begin{equation}\label{eq:gamma}
\Gamma^\textrm{plug}_{m}=\Gamma_m\times f\exp(-U^\ast/k_BT),
\end{equation}
where $U^\ast$ is the characteristic potential of the plug and $f$ is a factor depending on the details of the plug beam such as shape, position and power. We find this model nicely fit to our data with $C' T^{-2}\exp(-U^\ast/k_B T)+\Gamma_b$, where $U^\ast=k_B \times 25(1)~\mu$K and $C'=97(6)~\mu$K$^2$/s ($f\approx0.7$). This exponential behavior clearly supports our picture of the plugging effect. Note that $U^\ast$ is consistent with the deviation point in the $N$-$T$ evaporation trajectories and close to the potential at the trap bottom $U_0\approx k_b\times15~\mu$K. In our experiment, the collision rate was approximately 10$^3$ times faster than the loss rate, validating the thermal equilibrium assumption.

We characterized the heating effect of the Majorana loss with determination of the average loss energy of a spin-flipped atom $\varepsilon_{m} k_B T$. From the energy conservation relation $\dot E = - \varepsilon_{m}k_B T \dot N$, where $E=N \varepsilon k_B T$ is the total energy of the system with the average energy of a trapped atom $\varepsilon k_B T$, we derive a simple relation between the temperature and the atom number, $\dot T/T=(\varepsilon_{m}/\varepsilon -1)\dot N/N$. When $\varepsilon_{m}<\varepsilon$, the system is heated up and vice versa. In a power-law trap $U\propto r^{3/\delta}$, $\varepsilon=(3/2+\delta)$. Our measurements show that $\varepsilon_m$ has a very weak temperature dependence, slightly increasing at low temperature with the average value of 2.8, which is less than $\varepsilon$=4.5 for a linear trap ($\delta$=3) [Fig.~\ref{fig:Majorana_theory}(b)]. Remarkably, we observe that $\varepsilon_m$ is not affected by the optical plug in the temperature range of 10 to 100~$\mu$K. This is somewhat surprising because the average energy of a lost atom should increase in the presence of the plug due to the repulsive optical potential. We estimate that the effective increment in the average loss energy would be $U^\ast-U_0 \approx k_B\times 10~\mu$K with respect to the trap bottom and it might be observable at lower temperatures.

\section{Numerical simulation}\label{sec:numeric}

In this section, we numerically study the dynamics of evaporation process in an optically plugged quadrupole trap. Including the only three loss processes: background collision, rf evaporation and the Majorana spin-flip, the rate equations for atom number $N$ and temperatue $T$ are given as
\begin{subequations}\label{eq:lossrates}
\begin{eqnarray}
\frac{\dot N}{N} &=& -\Gamma_{b}  - \Gamma_{e} - \Gamma_{m} \\
\frac{\dot T}{T} &=& -(\frac{\varepsilon_e}{\varepsilon}-1)\Gamma_e - (
\frac{\varepsilon_m}{\varepsilon}-1)\Gamma_m,
\end{eqnarray}
\end{subequations}
where $\Gamma_e$ is the evaporation loss rate and $\varepsilon_e=\eta+\kappa$ is the average energy of an evaporated atom in $k_B T$. The first term in Eq.~(\ref{eq:lossrates}b) corresponds to the cooling from evaporation. We assume that the background collision loss occurs equally to all trapped atoms, thus not affecting the temperature.

The simulation for our experiment was performed in the following conditions: $N(0)$=$N_i$ and $T(0)$=$T_i$ at $t=0$~s, $\nu_\textrm{rf}(t)$=(30$-7t$/3)~MHz, $\eta$=$h \nu_\textrm{rf}/k_B T$, $\Gamma_b$=0.013/s and $\varepsilon_m$=0.28. The Majorana loss rate is parameterized with the plug beam power $P$ as $\Gamma_m$=$C(1-aP)T^{-2} \exp(-bP/T)$, where $C$=139~$\mu$K$^2/s$, $a$=0.176/W and $b$=1.48~$\mu$K/W. For 3D evaporation in a linear trap ($\delta$=3), standard evaporation theory~\cite{Luiten1996Evap} gives $\Gamma^\textrm{th}_e$=$n \sigma \bar{v} e^{-\eta} [\eta P(\frac{9}{2},\eta)-\frac{11}{2} P(\frac{11}{2},\eta)]/P(\frac{9}{2},\eta)^2$ and $\kappa$=$1-P(\frac{13}{2},\eta)/[\eta P(\frac{9}{2},\eta)-\frac{11}{2} P(\frac{11}{2},\eta)]$, respectively, where $P(a,\eta)$ is the incomplete gamma function~\cite{footnote4}. However, the observed evaporation speed was almost 2 times faster than the theory value. This is quite surprising because the theoretical estimation accounts for the maximum 3D evaporation, worthy of further investigation. We empirically set $\Gamma_e=1.95\times\Gamma^\textrm{th}_e$ in our simulation.

The simulated evaporation trajectories are presented with the experiment data in Fig.~\ref{fig:curve}. The trap decompression is not included in our simulation, so the trajectories deviate from the data points after 12~s. The maximum phase space density and the critical atom number $N_c$ at the phase transition were numerically estimated as a function of the plug beam power~[Fig.~\ref{fig:power}], which show a good agreement with the experiment results. We approximate the condensate number at 25\% of the critical atom number. The simulation with a perfect plug, i.e. $\Gamma_m$=0 (blue lines in Fig.~\ref{fig:curve}) suggests that the maximum condensate would have $N\approx2\times10^7$ atoms for the given 14~s evaporation.

\section{Conclusions}
We have described a rapid production of $^{23}$Na Bose-Einstein condensates in an optically plugged magnetic quadrupole trap. The off-axis optical plug generates a single global minimum in the combined trapping potential and consequently enhances the BEC production. The Majorana loss, which is critical in a quadrupole trap, has been quantitatively investigated, determining the average loss energy of a spin-flipped atom and the loss rate suppression of the optical plug. We have presented a simple model to explain the evaporation dynamics in the plugged quadrupole trap. This model will be helpful to develop and optimize evaporation processes using a magnetic quadrupole trap.

\section*{Acknowledgments}
The authors thank A. Keshet for experimental control software and M. Lee for experimental help. This work was supported by the WCU program (R32-2008-000-10045-0) and the NRF Grant (2010-0010172). J.C. acknowledges financial supports from MOE through BK21 fellowships.

\bibliographystyle{apsrev}

\begin{thebibliography}{15}
\bibitem{footnote1}
We could not achieve a condensate even with a 6~W plug beam when the beam was sent from a laser in free space. The laser was placed on a different optical table 4~m away from the machine.
\bibitem{footnote3}
The radial half width half maximum cloud size $l$ is approximately $2.5 (k_B T / \mu B')$ in an in-trap, column density distribution and $2\ln 2( k_B T/\mu B')$ in a 3D distribution.
\end{thebibliography}

\end{document}